# EFFECTS OF THE SPIN AND MAGNETIC MOMENTS ON THE INTERACTION ENERGY BETWEEN ELECTRONS. APPLICATION TO FERROMAGNETISM.

## V. DOLOCAN[1] AND V. O. DOLOCAN[2]

[1] *Faculty of Physics, University of Bucharest, Bucharest, Romania*
[2] *IM2NP & Aix-Marseille University, Faculté des Sciences et Technique, Avenue Escadrille Normandie Niemen, 13397 Marseille cedex 20, France*

**Abstract.**
We study the influence of the spin and magnetic moments on the energy of interaction between electrons. First, we separate the Heisenberg term and the second, we introduce the magnetic moment in the electron wave function. We have found that in an isolated atom, with an incompletely filled shell, the interaction energy for the triplet state is lower than that for the singlet state, so that the triplet state is favorable. Also, we have calculated the interaction energy between the electrons from the adjacent atoms in fcc and bcc lattices and we have found that the minimum interaction energy is attained for the triplet state.

Key words: ferromagnetism, electron-electron interaction, magnetic moment.

## 1. INTRODUCTION

An important problem is that of the origin of ferromagnetism. The physical effect that produces magnetic ordering of adjacent magnetic moments is the same that leads to magnetic ordering within an ion and produces Hund's rules. Interactions between spins are actually just a convenient way to record the end result of electrostatic repulsion[1]. Two electrons circling a nucleus reduce their Coulomb interaction by adopting an anti symmetric wave function that vanishes whenever they come near each other. The Pauli principle demands that the overall wave function must be antisymmetric, so the spin wave function must be symmetric. That is the electrons lower their energy by adopting the same spin state and developing a local magnetic moment. The important concepts are introduced by Heisenberg[2] and Stoner[3]. It was shown by Hubbard[4] that the correlation effects will lower the energy of non-magnetic states more than that of the ferromagnetic states, and so make the condition for ferromagnetism more stringent. The spin wave excitations give a way to explain the temperature dependence of the magnetization in magnetic materials at low temperatures[5].

In this paper, using an effective field Hamiltonian (detailed in Ref.6), we investigate the influence of the magnetic moments on the electron-electron interaction. We first discuss the separation of the Heisenberg term in the energy of interaction. Next, we present the modification of Coulomb's law by the interaction between the magnetic moments of the electrons. Further, we discuss the case of an isolated atom with applications to Hund's rules and the condition for ferromagnetism in *fcc* and *bcc* lattices.

## 2. MODEL AND FORMULATION

In this Section we present the electron-electron interaction by using an effective Hamiltonian[6] and use an way to separate the Heisenberg term. In the interaction picture the effective Hamiltonian of the electron-electron interaction is given by the expression (see Appendix):

$$H_{II}^{eff} = 2\hbar \sum_{q,q',k,k',\sigma,\sigma'} |g_{q_o}|^2 \frac{-1}{\varepsilon_k - \varepsilon_{k-q} - \omega_q} a_q^+ a_{q_o} a_{q_o}^+ a_q c_{k-q,\sigma}^+ c_{k',\sigma'}^+ c_{k,\sigma} c_{k'-q,\sigma'} \qquad (1)$$

$$g_{q_o} = \frac{\hbar D}{8N^2 mr(\rho_o + Dr/c^2)} \frac{qq_o}{\omega_q \omega_{q_o}} \sum_j e^{i*q_{oj}*r_j}$$

where $D$ is a coupling constant, $m$ is the mass of an electron, $r$ is the distance between the two electrons, $\rho_o$ is the massive density of the interacting field, $Dr/c^2$ is the "mass less density" of the interacting field, $\omega_q = cq$ is the classical oscillation frequency of the interacting field, $\omega_{qo}$ is the oscillation frequency of an electron, $q$ is the wave vector of the interacting field, $q_o$ is the wave vector of the boson associated with the electron, $k$ is the wave vector of the electron and $\varepsilon_k = \hbar k^2/2m$. Introducing creation and annihilation operators $c_{j\sigma}^+$ and $c_{j\sigma}$ of an electron of spin $\sigma$ in the orbital state $\phi(r - r_j)$ one can also write

$$c_{\mathbf{k-q},\sigma}^+ = \frac{1}{\sqrt{N}} \sum_j e^{-i(\mathbf{k-q})\cdot\mathbf{r}_j} c_{j\sigma}^+; \quad c_{\mathbf{k'},\sigma'}^+ = \frac{1}{\sqrt{N}} \sum_{j'} e^{-i\mathbf{k'}\cdot\mathbf{r}_{j'}} c_{j'\sigma'}^+$$

$$c_{\mathbf{k},\sigma} = \frac{1}{\sqrt{N}} \sum_{j'} e^{i\mathbf{k}\cdot\mathbf{r}_{j'}} c_{j'\sigma}; \quad c_{\mathbf{k'-q},\sigma'} = \frac{1}{\sqrt{N}} \sum_j e^{i(\mathbf{k'-q})\cdot\mathbf{r}_j} c_{j\sigma'}$$

and therefore,

$$c_{\mathbf{k-q},\sigma}^+ c_{\mathbf{k'},\sigma'}^+ c_{\mathbf{k},\sigma} c_{\mathbf{k'-q},\sigma'} = \sum_{jj'} c_{j\sigma}^+ c_{j'\sigma'}^+ c_{j'\sigma} c_{j\sigma'} \qquad (2)$$

where we have used that

$$\frac{1}{N} \sum_j e^{i(\mathbf{k-k'})\cdot\mathbf{r}_j} = \delta(\mathbf{k'-k})$$

We define the occupation number operator

$$n_j = \sum_\sigma c_{j\sigma}^+ c_{j\sigma}$$

and the spin operators

$$s_j^\alpha = \sum_{\sigma\sigma'} c_{j\sigma}^+ \left(\frac{\sigma_{\sigma\sigma'}^\alpha}{2}\right) c_{j\sigma'}$$

where $\sigma^\alpha$, $\alpha = x,y,z$ are the Pauli matrices. Writing the components out explicitly, one has

$$s^x = \frac{1}{2}(c_j^+ c_{j'} + c_{j'}^+ c_j); \quad s^y = \frac{1}{2} i(c_{j'}^+ c_j - c_j^+ c_{j'})$$

$$s^z = \frac{1}{2}(c_j^+ c_j - c_{j'}^+ c_{j'}); \quad s^+ = s^x + is^y = c_j^+ c_{j'}; \quad s^- = s^x - is^y = c_{j'}^+ c_j$$

With some algebra one obtains

$$\sum_{\sigma\sigma'} c_{j\sigma}^+ c_{j\sigma} c_{j'\sigma'}^+ c_{j'\sigma} = \frac{1}{2} n_j n_{j'} + 2s_j^z s_{j'}^z + s_j^+ s_{j'}^- + s_j^- s_{j'}^+ = \frac{1}{2} n_j n_{j'} + 2\mathbf{s}_j \cdot \mathbf{s}_{j'}$$

Therefore,

$$H_{I1}^{eff} = \sum_{jj'} K_{jj'} \left\{ \frac{1}{2} n_j n_{j'} + 2\mathbf{s}_j \cdot \mathbf{s}_{j'} \right\} \quad (3)$$

where

$$K_{jj'} = \frac{\hbar^3 D^2}{32 m^2 R^2 \left(\rho_o + \frac{DR}{c^2}\right)^2} \sum_{q,q_o,k} \frac{(q \cdot q_o)^2}{\omega_q^2 \omega_{q_o}^2} \frac{1}{2} \frac{e^{i(q_{oj} \cdot r_j - q_{oj'} \cdot r_{j'})}}{(\varepsilon_k - \varepsilon_{k-q}) - \omega_q} (n_q + 1)(n_{q_o} + 1) \quad (3a)$$

The first term enclosed between parentheses is a "pure" Coulomb term and the second term is the Heisenberg term

$$H_H = 2 \sum_{jj'} K_{jj'} \mathbf{s}_j \cdot \mathbf{s}_{j'} \quad (4)$$

## 3. EFFECT OF MAGNETIC MOMENT ON THE INTERACTION ENERGY.

Let us consider two electrons separated at a distance $r = |r_j - r_{j'}|$. The expectation value of the first term from Eq. (3) is

$$E_C = \sum_{jj'} K_{jj'} \frac{1}{2} n_j n_{j'} \quad (5)$$

For $n_q = n_{qo} = 0$ and $n_j = n_{j'} = 1$, we have

$$\frac{1}{2} \sum_{jj'} e^{i(q_{oj} \cdot r_j - q_{oj'} \cdot r_{j'})} = 1 + \cos\left[i(q_{o1} \cdot r_1 - q_{o2} \cdot r_2) - \frac{ie}{\hbar c}\left(\oint A(r_1) dr_1 - \oint A(r_2) dr_2\right)\right] = 1 + \cos(\Gamma) \quad (6)$$

where in the presence of the magnetic field we introduce the vector potential and thus we substitute $q_o \cdot r$ by $q_o \cdot r - (e/\hbar c) \oint A \cdot dr$.

The energy levels of atomic electrons are affected by the interaction between the electron spin magnetic moment and the orbital magnetic moment of the electron. It can be visualized as a magnetic field caused by the electron's orbital motion interacting with the spin magnetic moment. The effective magnetic field can be expressed in terms of the electron orbital angular momentum. We consider the vector potential $\mathbf{A} = \frac{\boldsymbol{\mu} \times \mathbf{r}}{r^3}$ where $\boldsymbol{\mu}$ is the magnetic dipole moment and $\mathbf{r}$ is a vector from the middle of the loop to an observation point. An electron in a stationary state in an atom, having a definite angular momentum projection $L_z = \hbar m_l$ ($m_l$ the quantum magnetic number), has a magnetic moment $\mu_z^{(l)} = \mu_B m_l$ where $\mu_B = e\hbar/2mc$ is the Bohr magneton. The theory and experiments demonstrate that the free electron has a magnetic moment equal to the Bohr magneton $\mu_B$, and a spin momentum $\mathbf{s}$, the projection of which on a specified direction are $s_z = \pm\hbar/2 = \hbar m_s$ where $m_s = \pm 1/2$ is the spin quantum number. For $\mu_z^{(s)} = \mu_B g m_s$ with $g = 2$ we obtain:

$$\Gamma = \frac{1}{2}(q_j + q_{j'})(r_j - r_{j'}) + \frac{1}{2}(q_j - q_{j'})(r_j + r_{j'}) - \frac{e}{\hbar}\frac{e^2}{4\pi mc^2}\frac{h}{e}\oint \frac{\mathbf{m}_{lj} \times \mathbf{r}_j}{r_j^3} \cdot dr_j +$$

$$\frac{e}{\hbar}\frac{e^2}{4\pi mc^2}\frac{h}{e}\oint \frac{\mathbf{m}_{lj'} \times \mathbf{r}_{j'}}{r_{j'}^3} \cdot dr_{j'} - \frac{e}{\hbar}\frac{e^2}{4\pi mc^2}\frac{h}{e}\oint \frac{2\mathbf{m}_{sj} \times \mathbf{r}_{jj'}}{r_{jj'}^3} dr_{jj'} + \frac{e}{\hbar}\frac{e^2}{4\pi mc^2}\frac{h}{e}\oint \frac{2\mathbf{m}_{sj'} \times \mathbf{r}_{jj'}}{r_{jj'}^3} \cdot dr_{jj'}$$

where $(h/e)\mathbf{m}_{li}$ and $(h/e)\mathbf{m}_{si}$ are the flux vectors. For $q_{oj} = q_{oj'} = q_o$, one obtains

$$\Gamma = q_o R \cos\theta - \Gamma_o$$

$$\Gamma_o = \frac{e^2}{2mc^2}\left[\frac{2\pi}{r_j}m_{lj} - \frac{2\pi}{r_{j'}}m_{lj'}\right] + \frac{e^2}{2mc^2}\left[-\frac{2\pi \times 2m_{sj}}{r_{jj'}} + \frac{2\pi \times 2m_{sj'}}{r_{j'j}}\right]$$

In these conditions $\mathbf{q}'_o = \mathbf{q}_o - \frac{e^2}{2m}\frac{(m_l + 2m_s)}{r^2}\hat{\mathbf{x}}$ where $\mathbf{x}$ is a unit vector perpendicular to $r$ and $\mu$. When the interacting field is a photon field, then $\rho_o = 0$. For a quasi free electron $\varepsilon_k - \varepsilon_{k-q} \ll \omega_q$, $\omega_{q_o} = q_o^2/2m$ ($m$ is an effective electron mass).

To calculate $K_{jj'}$ (3a) we consider the two sums separately. The first sum can be easily calculated as:

$$\sum_q \frac{(q \cdot q_o)^2}{\omega_q^3 \omega_{q_o}^2} = \left(\frac{2m}{\hbar}\right)^2 \frac{1}{q_o^2 c^3}\frac{\Omega}{(2\pi)^2}\int_0^\pi \cos^2\alpha \sin\alpha\, d\alpha \int_0^{q_o} q\, dq = \left(\frac{2m}{\hbar}\right)^2 \frac{r^3}{9\pi c^3}$$

where we have considered $\Omega = 4\pi r^3/3$. Next

$$\sum_{\mathbf{q}_o}(1 + \cos\Gamma) = 1 + \frac{\Omega}{(2\pi)^2}\int_0^{0.76\pi/r} q_o^2 dq_o \int_0^\pi \cos(q_o R \cos\theta - \Gamma_o)\sin\theta\, d\theta = 1 + 0.53\cos(\Gamma_o)$$

In these condition $E_C$ (5) becomes

$$E_C = \frac{\hbar c}{144\pi r}\left[1 + 0.53\cos(\Gamma_o)\right] \qquad (7)$$

The upper limit of $q_0$ as $0.76\pi/r$ is imposed by the constraint condition $[(4\pi r^3/3)/(2\pi)^3] \times 4\pi q_o^3/3 = 1$. For $m_{si} = m_{sj} = \frac{1}{2}$, $\Gamma_o$ reduces to:

$$\Gamma_o = \frac{\pi e^2}{mr_o c^2}\left(m_{lj}^\square - m_{lj'}\right) \qquad (8)$$

where $r_o = r_j = r_{j'}$ is the radius of the electron orbit and $r = r_{jj'}$. For $m_{lj} - m_{lj'} = 1$ we get:

$$\Gamma_{0^{triplet}} = \frac{\pi e^2}{mr_o c^2} = \frac{a}{r_o}$$

where $a = 8.95 \cdot 10^{-13}$ cm. (8a)

For $m_{sj} = \frac{1}{2}$, $m_{sj'} = -1/2$, we obtain:

$$\Gamma_0 = \frac{\pi e^2}{m r_o c^2}(m_{lj} - m_{lj'}) + \frac{2\pi e^2}{m r c^2} \tag{9}$$

For $m_{lj} = m_{lj'}$ the above relation reduces to:

$$\Gamma_{o^{singlet}} = \frac{2\pi e^2}{m r c^2} = \frac{2a}{r} \tag{9a}$$

For an interacting pair of spins $s_1 = s_2 = 1/2$, the expectation value of the Heisenberg term (4) is written

$$E_H = 2K_{12}\mathbf{s}_1.\mathbf{s}_2 = K_{12}[\mathbf{S}.\mathbf{S} - \mathbf{s}_1.\mathbf{s}_2 - \mathbf{s}_2.\mathbf{s}_1] = E_C\left[S(S+1) - \frac{3}{4} - \frac{3}{4}\right] \tag{10}$$

with $\mathbf{S} = \mathbf{s_1} + \mathbf{s_2}$. In the singlet state, $S = 0$ and $E_H = -(3/2)E_C$. In the triplet state, $S = 1$, and $E_H = (1/2)E_C$, where $E_C$ is given by Eq. (7). The spin dependent energy of interaction is

$$E_S = \frac{\hbar c}{r}\frac{0.53\cos(\Gamma_o)}{144\pi} + \frac{\hbar c}{r}\frac{1 + 0.53\cos(\Gamma_o)}{144\pi}\left[S(S+1) - \frac{3}{2}\right] \tag{11}$$

The term responsible for magnetic ordering is that containing the product of $\cos(\Gamma_o)$ and $S(S+1)$, that is

$$E_M = \frac{0.53\hbar c\cos(\Gamma_o)}{144\pi r}S(S+1) \tag{12}$$

For $S = 0$, singlet state, one obtains $E_M^S = 0$, while for $S = 1$, triplet state, one obtains

$$E_M^t = \frac{1.06\hbar c\cos(\Gamma_o^t)}{144\pi r} \tag{13}$$

It is observed that for $\cos(\Gamma_o) < 0$, we have $E_M^t < E_M^s$ and the interaction energy for the triplet state is lower than that associated with the singlet state. But, in this case, as it appears from relation (9a), the distance between the two electrons should be $r < 5.7 \cdot 10^{-13}$ cm. This situation may occur in a shell of an atom or in an electron band, because the *cosine* term diminish the repulsion Coulomb energy. Thus for an atom with two valence electrons in a shell with $l > 1$, the interaction energy between the two electrons when they occupy different $L_z$ states (all degenerate in energy) and adopt a triplet spin state, is smaller than the interaction energy when they occupy the same $L_z$ state, in which case they must have opposite spins. Therefore, the triplet state is favorable. This may explain Hund's rules related to the organization of electrons in incompletely filled shells. Finally, we note that if we do not separate the Heisenberg term, and do not take into account the magnetic moments, the expectation value of $H_I$ (3) would be[6]

$$E_I = \frac{1}{147}\frac{\hbar c}{r} \approx \alpha\frac{\hbar c}{r} \tag{14}$$

which is just the Coulomb's law where the constant 1/147 differs some few on the fine structure constant, $\alpha = 1/137$; this difference appears due to constraint in the evaluation of the integral over $q_o$. For a better agreement with experimental data we must take in the above equatuions the value of 0.65 than 0.53.

## 3. RESULTS FOR FCC AND BCC LATTICES

Suppose now that at some instant a given atom has its total spin in the up direction. The intra-atomic interactions are of such a nature that this atom tends to attract electrons with spin up and repel those with spin down. In the case of the face-centered cubic lattice every atom is surrounded by 12 nearest neighbors, which have the coordinates $(a/2)(\pm 1,\pm 1,0)$, $(a/2)(\pm 1,0,\pm 1)$, $(a/2)(0,\pm 1,\pm 1)$. Suppose the magnetic moment $\boldsymbol{\mu}(0,0,\mu)$. Then, the vector potential can be chosen in the following form $\mathbf{A}(A_x,A_y,0)$. For $n_j = n_{j'} = 1$, $2\mathbf{s}_j \cdot \mathbf{s}_{j'} = S(S+1) - 3/2$ we write:

$$\frac{1}{N}\sum_{jj'} e^{i(q_j \cdot r_j - q_{j'} \cdot r_{j'})} = 4\left[3 - \cos\left(\frac{aq_{ox}}{2} - \Gamma_o\right)\cos\left(\frac{aq_{oy}}{2} - \Gamma_o\right) - \cos\left(\frac{aq_{ox}}{2} - \Gamma_o\right)\cos\left(\frac{aq_{oz}}{2}\right) - \cos\left(\frac{aq_{oy}}{2} - \Gamma_o\right)\cos\left(\frac{aq_{oz}}{2}\right)\right] \quad (15)$$

where $\Gamma_o$ is given by Eq. (8a) for the triplet state and by Eq.(9a) for the singlet state. The sum in the interaction energy becomes:

$$\frac{1}{N^3}\sum_{q,q'}\frac{(q \cdot q'_o)^2}{\omega_q^3 \omega_{q'}^2}\sum_{jj'} e^{i(q_{oj} \cdot r_j - q_{oj'} \cdot r_{j'})} = \left(\frac{2m}{\hbar}\right)^2 \frac{r^3}{9\pi c^3}\frac{\Omega}{(2\pi)^3} 4 \int_0^{\pi/a} dq'_{ox} \int_0^{\pi/a} dq'_{oy} \int_0^{\pi/a} dq'_{oz}$$

$$\left[3 - \cos\left(\frac{aq'_{ox}}{2} - \Gamma_o\right)\cos\left(\frac{aq'_{oy}}{2} - \Gamma_o\right) - \cos\left(\frac{aq'_{ox}}{2} - \Gamma_o\right)\cos\left(\frac{aq'_{oz}}{2}\right) - \cos\left(\frac{aq'_{oy}}{2} - \Gamma_o\right)\cos\left(\frac{aq'_{oz}}{2}\right)\right]$$

$$\frac{m^2}{\hbar^2}\frac{64 r^3}{9\pi^3 c^3}\left[3\pi^2 - f_{fcc}(\Gamma_o)\right]$$

where

$$f_{fcc}(\Gamma_o) = (\cos\Gamma_o + \sin\Gamma_o)^2 + 2(\cos\Gamma_o + \sin\Gamma_o)$$

Further

$$E_C = \sum_{jj'} K_{jj'} = \frac{\hbar c}{9\pi^3 r}\left[\frac{3\pi^2}{4} - (\cos\Gamma_o + \sin\Gamma_o)^2 - 2(\cos\Gamma_o + \sin\Gamma_o)\right]$$

and the interaction energy

$$E_I = \frac{\hbar c}{9\pi^3 r}\left[\frac{3\pi^2}{4} - (\cos\Gamma_o + \sin\Gamma_o)^2 - 2(\cos\Gamma_o + \sin\Gamma_o)\right]\left\{\frac{1}{2} + S(S+1) - \frac{3}{2}\right\} \quad (16)$$

The term responsible for magnetic ordering is

$$E_M = -\frac{\hbar c \left[\left(\cos\Gamma_o + \sin\Gamma_o\right)^2 + 2\left(\cos\Gamma_o + \sin\Gamma_o\right)\right]}{9\pi^3 r} S(S+1) = -\Delta E_I S(S+1) \qquad (16a)$$

which has a minimum for $\Gamma_o = 2n\pi - \pi/4$.

In a *bcc* structure, every atom is surrounded by 8 nearest neighbors which have the coordinates $(a/2)(\pm 1, \pm 1, \pm 1)$. One obtains

$$\frac{1}{N}\sum_{jj'} e^{i(\mathbf{q}_{oj}\cdot\mathbf{r}_j - \mathbf{q}_{oj'}\cdot\mathbf{r}_{j'})} = 8\left[1 - \cos\left(\frac{aq_{ox}}{2} - \Gamma_o\right)\cos\left(\frac{aq_{oy}}{2} - \Gamma_o\right)\cos\left(\frac{aq_{oz}}{2}\right)\right]$$

Further,

$$\frac{1}{N^3}\sum_{q,q_o}\frac{(q\cdot q_o)^2}{\omega_q^3 \omega_{q_o}^2}\sum_{jj'} e^{i(q_{oj}\cdot r_j - q_{oj'}\cdot r_{j'})} = \frac{4m^2}{\hbar^2}\frac{r^3}{9\pi c^3}\frac{\Omega}{\pi^3}$$

$$8\int_0^{\pi/a} dq_{ox} \int_0^{\pi/a} dq_{oy} \int_0^{\pi/a} dq_{oz}\left[1 - \cos\left(\frac{aq_{ox}}{2} - \Gamma_o\right)\cos\left(\frac{aq_{oy}}{2} - \Gamma_o\right)\cos\frac{aq_{oz}}{2}\right] =$$

$$\frac{32 m^2}{\hbar^2}\frac{r^3}{9\pi c^3}\left[1 - \frac{8}{\pi^3}\left(\cos\Gamma_o + \sin\Gamma_o\right)^2\right]$$

The energy of interaction is given by the expression

$$E_I = \frac{\hbar c}{18\pi r}\left[1 - \frac{8}{\pi^3}\left(\cos\Gamma_o + \sin\Gamma_o\right)^2\right]\left[\frac{1}{2} + S(S+1) - \frac{3}{2}\right]$$

and the term responsible for magnetic ordering is

$$E_M = -\frac{4\hbar c}{9\pi^4 r} f_{bcc} S(S+1) = -\Delta E_I S(S+1)$$
$$f_{bcc} = \left(\cos\Gamma_o + \sin\Gamma_o\right)^2 \qquad (17)$$

which is some few different from the energy (16a) obtained for the *fcc* structure.
From Eqs. (16) and (17) it appears that there is an oscillating interaction between the two electrons belonging to different atoms. The nature of this interaction may be not different than that which is encountered in the case of interaction between two ions by indirect exchange [7]. In the later case, roughly speaking, the electrons belonging to an ion flip a conduction electron which then travels to another site and interacts with the spin of the ion of the second site. Perhaps the most significant application of the RKKY theory has been to the theory of giant magneto resistance (GMR)[8,9]. GMR was discovered when the coupling between thin layers of magnetic materials separated by a non-magnetic spacer material was found to oscillate between ferromagnetic and anti ferromagnetic as a function of the distance between the layers. The period of an oscillation is determined by the Fermi wave vector, in the case of free electron gas, via $\lambda = 2\pi/2k_F = \pi/k_F$. The phase of the *cosine* in RKKY theory is $k_F r$ while in our theory this phase is $\Gamma_o = \pi e^2/mrc^2 = 1/k_o r$ where $k_o = mc^2/\pi e^2 = 1.296.10^{12}$ cm$^{-1}$. In this case the "period" of an oscillation is dependent of $r$. From the condition $\Gamma_o^{(n)} - \Gamma_o^{(n+1)} = 2\pi$, one obtains $\lambda_n = r_n - r_{n+1} = 2k_o r_n r_{n+1}$. For $r_n \sim 10^{-5}$ Å and $m = m_o$, results $\lambda_n \sim 10^{-5}$ Å a value which is much shorter than the Compton wavelength, $\hbar/mc = 8\times 10^{-3}$ Å, and than the period of the RKKY oscillation, which is of the order of 3Å. The oscillatory behavior is the result of the interference of two oscillating fields generated by the magnetic moments of the two electrons. Each fringe arises from a definite difference in phase. In the RKKY model the effect is an oscillatory polarization of the conduction electrons. Due to the small values of parameter *a*, for the distance between electrons of ~ 3 Å in a lattice, it results $\Gamma_o \ll 1$ and the effect of *cosine* terms is negligible.

## 4. CONCLUSIONS

Using an effective field Hamiltonian (as in Ref.6) we have shown that the Coulomb's law is modified by the spin-spin interaction. First, we separate the Heisenberg term in the energy of interaction. Next, we take into account the effect of the spin magnetic moment. The Coulomb interaction is modulated by a *cosine* term whose argument depends on the spin magnetic moment and on the distance between the two electrons. Evidently, this situation occurs in the absence of the magnetic field, for example in superconductors. Further, we have studied the influence of the electron magnetic moments, both orbital and spin, on the electron-electron interaction in an isolated atom. We have found that in an incomplete shell the interaction energy of the triplet state is smaller than the energy of the singlet state. On this basis may be explained Hund's rules. The condition of ferromagnetism is studied in both *fcc* and *bcc* lattices.. The coupling energy, due to the magnetic field generated by the two interacting electrons, oscillate with a "period" $1/k_o r$ where $k_o = mc^2/\pi e^2$. This period of oscillation is dependent of *r* and is of the order of $10^{-5}$ Å, a value which is much smaller than the RKKY period which is of ~ 3Å.

## APPENDIX

The electron will always carry with it a lattice polarization field. The composite particle, electron plus phonon field, is called a polaron; it has a larger effective mass than the electron in the unperturbed lattice. By analogy, in our model, we consider a coupling between an electron and a boson.

Let us consider a linear chain of N bodies, separated at a distance R. The Hamiltonian operator of the interacting bodies ( electrons ) and the boson connecting field takes the general form

$$H = H_{o,el} + H_{o,ph} + H_I$$

where

$$H_{o,el} = \sum_{k,\sigma} \frac{\hbar^2 k^2}{2m} c^+_{k,\sigma} c_{k,\sigma}$$

is the Hamiltonian of the electrons of mass *m*, $c^+_{k,\sigma}, c_{k,\sigma}$ are the electron creation and annihilation operators, *k* is the wave vector of an electron and σ ls the spin quantum number,

$$H_{o,ph} = \sum_q \hbar \omega_q \left( a^+_q a_q + \frac{1}{2} \right)$$

$$\omega_q = \left( \frac{\alpha + DRq^2}{\rho} \right)^{1/2}$$

where $\omega_q$ is the classical oscillation frequency, $\alpha$ is the restoring force constant, D is the coupling constant, $\rho$ is the linear density of flux lines, $a_q^+$, $a_q$ are the boson creation and annihilation operators and $\rho = \rho_o + DR/c^2$, $\rho_o$ is the density of the interacting field, if this is a massive field, $c$ is the velocity of the boson waves. The interaction Hamiltonian operator $H_I$ is given by the expression[10]

$$H_I = \frac{DR}{2} \int \sum_n s_n \left(\frac{\partial \mathbf{u}}{\partial \mathbf{z}}\right)^2 \Psi^+(\mathbf{z})\Psi(\mathbf{z}) d\mathbf{z}$$

where

$$u(z) = \frac{1}{\sqrt{NR}} \sum_{q,l} e_{ql} \left(\frac{\hbar}{2\rho\omega_{ql}}\right)^{1/2} \left(a_{ql} e^{iq.z} + a_{ql}^+ e^{-iq.z}\right)$$

$$\Psi(z) = \frac{1}{\sqrt{NR}} \sum_{k,\sigma} c_{k\sigma} e^{ik.z} \chi(\sigma)$$

$$\Psi^+(z) = \frac{1}{\sqrt{NR}} \sum_{k,\sigma} c_{k\sigma}^+ e^{-ik.z} \chi(\sigma)$$

$$s_n = \frac{1}{N} \sum_{q,q_o} S_q e^{iq_o(z-z_n)}; \quad S_q = \frac{1}{R} \frac{\hbar}{2m\omega_q} \left(b_{q_o} + b_{-q_o}^+\right)$$

$b_{q_o}^+$, $b_{q_o}$ are creation and annihilation operators associated with the electron oscillations, $\mathbf{e}_{ql}$ denotes the polarization vectors and $\chi(\sigma)$ is the spin wave function. $s_n$ is the displacement of a body near its equilibrium position in the direction of $\mathbf{R}$ and, in the approximation of nearest neighbours, it is assumed that $D$ does not depend on $n$. The Hamiltonian of interaction between electrons via bosons becomes

$$H_I = -\frac{1}{N^3} \frac{D}{2R^2} \int \sum_n \sum_{k,k',q,q',q_o,\sigma,\sigma'} \frac{\hbar}{2m\omega_{q_o}} \left(b_{q_o} + b_{-q_o}^+\right) \frac{\hbar q.q'}{2\rho\omega_q}$$

$$e^{iq_o.z_n} \left(a_q + a_{-q}^+\right)\left(a_{q'} + a_{-q'}^+\right) c_{k'\sigma'}^+ c_{k\sigma} e^{-i(q+q'+k-k'-q_o).z} dz$$

$\mathbf{q}$, $\mathbf{q'}$ are the wave vectors associated with the bosons of the connecting field, $\mathbf{q}_o$ is the wave vector associated with the oscillations of the electron, and $\mathbf{k}$, $\mathbf{k'}$ are the wave vectors of the electrons. Consider the integral over z

$$\int e^{-i(\mathbf{q}+\mathbf{q'}+\mathbf{k}-\mathbf{k'}+\mathbf{q_o}).\mathbf{z}} dz = NR\Delta\left(\mathbf{q}+\mathbf{q'}+\mathbf{k}-\mathbf{k'}-\mathbf{q_o}\right) \qquad (A1)$$

where $\Delta(x) = 1$ for $x = 0$ and $\Delta(x) = 0$, otherwise. In the bulk crystal NR is replaced by $V = N\Omega$ where $\Omega$ is the volume of a unit cell and N is the number of unit cells. We write

$$H_I = -\hbar \sum_{k,k',q,q',q_o,\sigma,\sigma'} g_{q_o} \left(a_q + a_{-q}^+\right)\left(a_{-q'} + a_{q'}^+\right) c_{k'\sigma'}^+ c_{k\sigma} b_{q_o} \Delta\left(q+q'+k+k'-q_o\right) +$$

$$g_{q_o} \left(a_{-q} + a_q^+\right)\left(b_{q_o} + b_{-q_o}^+\right) c_{k'\sigma'}^+ c_{k\sigma} b_{q_o}^+ \Delta\left(q+q'+k-k'+q_o\right)$$

where

$$g_{q_o} = \frac{\hbar D}{8N^2 mR(\rho_o + DR/c^2)} \frac{qq'}{\omega_q \omega_{q'}} \sum_n e^{iq_o \cdot z_n}$$

If we omit the terms with two creators and two annihilators, it may be written

$$(a_\mathbf{q} + a^+_{-\mathbf{q}})(a_{-\mathbf{q'}} + a^+_{\mathbf{q'}}) \approx a_\mathbf{q} a^+_{\mathbf{q'}} + a^+_{-\mathbf{q}} a_{-\mathbf{q'}}$$

In equation (A1) we choose **q' =q$_o$, k' = k + q.** In the interaction picture the effective Hamiltonian is given by

$$H_I^{eff} = H_{I1}^{eff} + H_{I2}^{eff}$$

$$H_{I1}^{eff} = \hbar \sum_{q,q_o,k} |g_{q_o}|^2 \frac{\omega_q}{(\varepsilon_k - \varepsilon_{k-q})^2 - \omega_q^2}$$

$$\left( a_q a^+_{q_o} a_{q'} a^+_{q'} + a^+_q a_{q_o} a^+_{q'} a_{q'} \right) c^+_{k-q,\sigma} c^+_{k'\sigma'} c_{k\sigma} c_{k'-q,\sigma'}$$

$$H_{I2}^{eff} = 2\hbar \sum_{q,k} |g_q|^2 \frac{1}{(\varepsilon_k - \varepsilon_{k-q}) - \omega_q} a^+_q a_q c^+_{k-q,\sigma} c_{k-q,\sigma}$$

The expectation value of the energy of $H_{I1}^{eff}$ is the energy of the electron-electron interaction given in the text. The expectation value of the energy of $H_{I2}^{eff}$ is the self energy of the electron and is used to calculate, for example, the polaron energy[6]. Finally, we specify that we have considered the interaction, via bosons, of two oscillators which move with the momentum $\hbar k$. The result is also valid for *k* = 0.

**REFERENCES**
   1. W. Heitler and F. London, Z. Phys. 44, 455(1927).
   2. W. Heisenberg, On the theory of ferromagnetism, Z. Phys. 49, 619(1926).
   3. E. C. Stoner, *Magnetism and Matter*, (Methuen, Lodon, 1934).
   4. J. Hubbard, Proc. Roy. Soc. of London A266, 278(1963).
   5. K. Hüller, J. Mag. Magn. Mater. 61, 347(1986)
   6. V. Dolocan, A. Dolocan, and V. O. Dolocan, Int. J. Mod. Phys. B 24, 479(2010); *ibid* Rom. J. Phys. 55, 153 (2010).
   7. C. Kittel, Indirect exchange interaction in metals, *Solid State Physics: Advances in Research and Applications*, 22, 1(1968).
   8. P. Grünberg, R. Schreiber, Y. Pang, M. B. Brodsky, and H. Sowers, Phys. Rev. Lett. 57, 2442(1986).
   9. M. N. Baibich, J. M. Broto, A. Fert, F. Nguyen van Dan, F. Petroff, P. Etienne, G. Creuzet, A. Friederich, and J. Chazelos, Phys. Rev. Lett. 61, 2472(1988)
   10. A. Dolocan, V. O. Dolocan and V. Dolocan, Mod. Phys. Lett. 19, 669(2005).

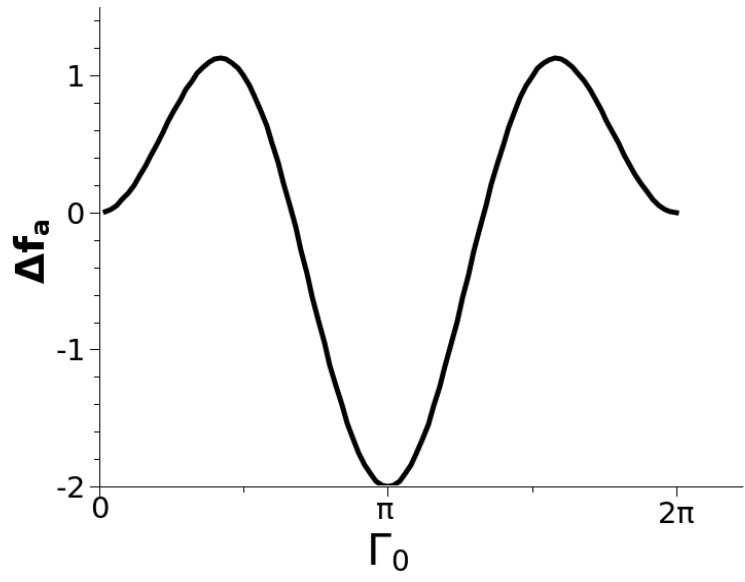

Fig.1